\documentstyle[12pt]{article}
\def\be{\begin{eqnarray}}
\def\ee{\end{eqnarray}}
\def\ba{\begin{array}}
\def\ea{\end{array}}

\begin{document}

\begin{center}
{\bf\Large {Absolute Time and Temperature in Quantum and \vskip 3mm
Classical Relativistic Mechanics}}
\end{center}
\vskip 1cm
\begin{center}
{\bf \large {Vadim V. Asadov$^{\star}$\footnote{asadov@neurok.ru}
and Oleg V.
Kechkin$^{+\,\star}$}\footnote{kechkin@depni.sinp.msu.ru}}
\end{center}

\vskip 5mm

\begin{center}
$^+$Institute of Nuclear Physics,\\
Lomonosov Moscow State University, \\
Vorob'jovy Gory, 119899 Moscow, Russia
\end{center}

\vskip 5mm

\begin{center}
$^\star$ Neur\,OK--III,\\
Scientific park of MSU, Center for Informational Technologies--104,\\
Vorob'jovy Gory, 119899 Moscow, Russia
\end{center}

\vskip 1cm

\begin{abstract}
We present a relativistic quantum mechanics of a point mass with
absolute thermodynamic time and temperature, combined to a single
complex parameter of evolution. In this theory, the geometric time
is introduced as one of space-time coordinates; it does not coincide
with the thermodynamic time on the kinematical level. It is
established, that the theory allows a consistent dynamics with
nontrivial probability density at the limit $\hbar\rightarrow 0$. We
identify this dynamics with the classical one, and prove its
relativistic invariance. It is shown, that the thermodynamical time
becomes proportional to the proper time of the point mass in the
conventional classical regime of its evolution with the
delta-functional distribution of the probability density.
\end{abstract}

\vskip 0.5cm

PACS No(s).\, : 05.30.-d,\,\,05.70.-a.

\vskip 1cm

One of the most intriguing problems of the modern theoretical
physics is related to contradiction between invertible character of
all known fundamental dynamical theories and unidirectional
evolution of the real world, which is governed by the second law of
thermodynamics. The simplest and the most natural way to improve
this strange situation is to develop fundamental `microtheories'\,
with well-defined arrow of time, and after that to ground the
thermodynamics on the base of their irreversible dynamics. We think,
that the phenomena of arising of the arrow of time
(\cite{atf}--\cite{atl}) is fundamental itself. It cannot be
generated by any procedure of average out type in framework of some
`macrotheory'\, using a really consistent mathematics.

Recently we have shown, that quantum theory with complex parameter
of evolution and non-Hermitian Hamiltonian structure provides an
adequate framework for description of quantum systems with the such
unidirectional dynamics. In this theory, the `usual'\, time is the
real part of the parameter of evolution, whereas the corresponding
imaginary part is proportional to the inverse absolute temperature
of the system. Then, the Hermitian part of the Hamiltonian is
identified with the energy operator of the theory, and the
anti-Hermitian one represents its decay operator. This generalized
quantum theory is considered as analytical in respect to the complex
time; the decay operator is taken as commuting one with the operator
of energy \cite{q-1}.

Then, it was demonstrated, how one can define the essentially
macroscopic thermodynamic regime of the evolution. For the
isothermal and adiabatic regimes, it was proven the presence of the
arrow of time in dynamics of the quantum systems in the case of
nontrivial operator of decay. Also, it was presented one important
realization of this general theoretical scheme, which relates the
arrow of time in the dynamics with the droningly increasing parity
violation for the solutions \cite{q-2}. We think, that the dynamical
mechanism of the parity violation (not of the `left-right type'\,
only) has the universal sense, and can be the fundamental reason for
the validity of the second law of thermodynamics.

It was established a consistent classical limit of the quantum
theory under consideration. The corresponding dynamics is defined as
the quantum one in the limit $\hbar\rightarrow 0$. In this dynamics,
one defines a trajectory of maximum of the probability density,
which is not only the delta-distribution in the our variant of the
classical theory. It was proven, that this classical dynamics is
time-irreversible too (at least, for the corresponding regimes of
quantum evolution with arrow of time) \cite{q-c}. Thus, it is
possible to say about unification performed for the both quantum and
classical theories with the irreversible principles of
thermodynamics.

The only problem arisen in this approach is related to the very
special role of time in its formulation. However, this situation is
not unusual for the dynamics, represented in the Hamilton's form. Of
course, one can try to reformulate the theory in the Lagrange terms,
but it seems really problematic here. In this work we choose another
way to construct the transparently relativistic theory: we suggest
to use two different times -- the `thermodynamic'\, and
`geometric'\, ones. Here, the geometric time is understood as one of
the space-time coordinates; the Hamiltonian must be taken as
Poincare-invariant quantity to deal with the transparently
relativistic quantum theory.

In the quantum realization of the theory, the wave-function $\Psi$
depends analytically on the complex parameter of evolution $\tau$,
and also on the set of space-time coordinates $x^{\mu}$ (where
$\mu=0, ...,3$). Here
\be\label{rel1}\tau=t-i\,\frac{\hbar}{2}\,\beta,\ee where $t$ is the
thermodynamic time, and $\beta$ is proportional to the inverse
absolute temperature $T$ of the system ($\beta=1/kT$, where $k$ is
the Bolzman constant). We would like to stress again, that the
geometric time $x^0$ is not identic to the thermodynamic time $t$ --
these are the completely independent quantities. The Schrodinger's
equation and the analytical condition, imposed on the wave-function
$\Psi=\Psi(\tau, x^{\mu})$, read: \be\label{rel2}
i\hbar\Psi_{,\,\tau}={\cal H}\Psi, \qquad \Psi_{,\,\tau^*}=0.\ee
Here, the only non-specified quantity is the Hamiltonian ${\cal H}$,
which must be constructed using the energy operator $E$ and the
decay operator $\Gamma$ according to the relation ${\cal H}
=E-i\,{\hbar}\,\Gamma/{2}$ (see \cite{q-1}). In this work we do not
plan to study arrow of time effects in the theory, so we put
$\Gamma=0$. Then, to write down the energy operator in the
relativistic invariant form, one can start from the Einstein's
formula $E=mc^2$ for the system, describing a point particle with
the mass $m$. Let us consider this system, and rewrite the
`Einsten's'\, operator of energy in the Poincare-invariant form,
using the 4-moment $P_{\mu}$ and the Minkowskian metric $g_{0\,
\mu\nu}$ (with the signature $-+++$). The corresponding Hamiltonian
reads: \be\label{rel3} {\cal
H}=-\frac{1}{m}\,g_0^{\mu\nu}P_{\mu}P_{\nu}.\ee The last step is to
put \be\label{rel4} P_{\mu}=-i\,\hbar\,\partial_{\mu};\ee the
resulting theory (\ref{rel2})--(\ref{rel4}) has explicitly
relativistic invariance if one defines the thermodynamic quantities
$t$ and $\beta$ as the relativistic scalars. In doing so, we mean
the presence of some `universal thermostat',\, which can be
identified with the physical vacuum of the theory. Note, that the
thermodynamic and geometric times have completely different
transformation properties.

To construct the classical realization of this system, one must
represent the wave-function in the standard exponential form,
\be\label{rel5}\Psi=\exp{\left ( \frac{i}{\hbar}\,{\cal S}\right
)}.\ee Then, the complex phase ${\cal S}$ must be parameterized as
\be\label{rel6}{\cal S}={\cal S}_1+i\,\frac{\hbar}{2}\,{\cal S}_2\ee
in terms of the real functions ${\cal S}_1$ and ${\cal S}_2$ of the
real variables $t$, $\beta$, and $x^{\mu}$. This parametrization
allows one to save a nontrivial probability structure in the
classical limit of the theory \cite{q-c}. Actually, the quantity
$w=|\Psi|^2=\exp{(-{\cal S}_2)}$ defines the probability density
$\rho$ accordingly the relation $\rho=w/Z$, where $Z=\int dx\, w$.
Let us define the classical value of the given quantity as its limit
at $\hbar\rightarrow 0$. It is clear, that the classical value of
the probability density is defined by the classical value of the
function ${\cal S}_2$ completely. This means, that the nontrivial
probability can be a fundamental property of the classical systems
(i.e., of the $\hbar$-independent ones), if their dynamics supports
the essentially distributed character of the function ${\cal S}_2$.
Of course, in the case of the $\delta$-functional distribution, one
comes back to the standard deterministic variant of the classical
theory.

To establish the new classical formalism, let us rewrite, at the
first time, the quantum dynamical system (\ref{rel2}) in terms of
the real variables $t$ and $\beta$, using Eq. (\ref{rel1}). After
that, it is necessary to substitute the parametrization
(\ref{rel5})--(\ref{rel6}) into the dynamical equations, obtained at
the previous step. Finally, one must calculate the limit at
$\hbar\rightarrow 0$ for the four real dynamical equations of the
theory (they define all partial derivatives of the first order for
the quantities ${\cal S}_1$ and ${\cal S}_2$ in respect to the
variables $t$ and $\beta$). Let us denote the classical value of
${\cal S}_{\alpha}$ as ${\rm S}_{\alpha}$ ($\alpha =1,2$). Then, the
corresponding pair of the dynamical equations for the phase ${\rm
S}_1$ reads: \be\label{rel7}{\rm S}_{1,\,t}=-E,\qquad
\label{rel8}{\rm S}_{1,\,\beta}=0,\ee where now the energy function
has its classical value (in the conventional Hamilton-Jacobi sense):
\be\label{rel9}E=-\frac{1}{m}\,\,g_0^{\mu\nu}\,{\rm S}_{1,\,\mu}{\rm
S}_{1,\,\nu}.\ee  Thus, the function ${\rm S}_{1}$ does not depend
on the temperature. Moreover, its $t$-dependence is defined by the
standard Hamilton-Jacobi equation. Then, by setting
\be\label{rel10}{\rm S}_2={\rm S}_e+\beta\,E\ee on the classical
values of the all  functions written, one obtains for ${\rm S}_e$
the following defining system of equations: \be\label{rel11}{\rm
S}_{e,\,t}-\frac{2}{m}\,g_0^{\mu\nu}\,{\rm S}_{1,\,\mu}{\rm
S}_{e,\,\nu}=-\frac{2}{m}\,g_0^{\mu\nu}\,{\rm S}_{1,\,\mu\nu},\qquad
{\rm S}_{e,\,\beta}=0.\ee Thus, the `entropy function'\, ${\rm
S}_e={\rm S}_e(t,\,x^{\mu})$ is $\beta$-independent too; its
$t$-dependence is defined completely by the `action function'\,
${\rm S}_1={\rm S}_1(t,\,x^{\mu})$. Note, that the modified
classical dynamics under construction is the Cauchy problem with the
equations (\ref{rel7}), (\ref{rel11}), and the initial conditions
${\rm S}_{1}(\,t_0, x^{\mu})={\rm S}_{1\,0}( x^{\mu}), \,\, {\rm
S}_{e}(\,t_0, x^{\mu})={\rm S}_{e\,0}( x^{\mu})$ with the arbitrary
initial data ${\rm S}_{1\,0}(x^{\mu})$ and ${\rm S}_{e\,0}(x^{\mu})$
taken. The only restrictions on these data are related to the
existence of all integral quantities, which must be written in their
terms.

We define the classical trajectory of the point mass $m$ using the
necessary extremum condition for the probability density. It is
equivalent to the one for the function ${\rm S}_2$, calculated in
the classical limit. The necessary extremum condition, i.e. the
relation ${\partial {\rm S}_2}/{\partial x^{\mu}}=0$, defines the
classical world sheet $x^{\mu}=x^{\mu}(t,\,\beta)$ of the theory.
One can differentiate this relation in respect to $t$ and $\beta$ on
the world sheet, and obtain the following equations:
\be\label{rel14} x^{\mu}_{,\,t}=\frac{2}{m}\left [
-g_0^{\mu\nu}p_{\nu}+\left ( A_2^{-1}\right )^{\mu\nu}\Box {\rm
S}_{1,\,\nu} \right ],\qquad x^{\mu}_{,\,\beta}=\frac{2}{m}\left (
A_2^{-1}A_1\right )^{\mu}_{\,\,\nu}g_0^{\nu\lambda}p_{\lambda}.\ee
Here $\rm (A_{\alpha})_{\mu\nu}=\partial^2 {\rm S}_{\alpha}/\partial
x^{\mu}\partial x^{\nu}$\, ($\alpha=1,2$), and the introduced
momentum variables are defined as $p_{\mu}={\partial {\rm
S}_1}/{\partial x^{\mu}}$ (the derivatives are calculated on the
classical world sheet). Then, the dynamical equations for the
momentums read: \be\label{rel15} p_{\mu,\,t}=\frac{2}{m} \left (
A_1A_2^{-1}\right )_{\mu}^{\,\,\nu}\Box {\rm S}_{1,\,\nu},\qquad
p_{\mu,\,\beta}=\frac{2}{m}\left ( A_1A_2^{-1}A_1\right
)_{\mu\nu}g_0^{\nu\lambda}p_{\lambda}.\ee It is seen, that the
Hamilton-like system of the equations (\ref{rel14})--(\ref{rel15})
is transparently relativistic-invariant.

To detect the total $t$-dependence of the coordinates and moments,
which is defined by the total derivatives $\dot
x^{\mu}=x^{\mu}_{,t}+\dot\beta x^{\mu}_{,\beta}$ and $\dot
p_{\mu}=p_{\mu,t}+\dot\beta p_{\mu,\,\beta}$, one must fix the
`temperature curve'\, $\beta=\beta(t)$. However, in the `hard'\,
classical regime with the highly-resonance type of the probability
density, one obtains a great simplification for the dynamics under
consideration. In concrete situation with a (quasi) Gaussian
resonance, the main part of the `phase'\, ${\rm S}_2$ is given by
its second differential $d^2 {\rm S}_{2}={\rm
A}_{2\,\mu\nu}dx^{\mu}dx^{\nu}$ with the center on the classical
trajectory. It is clear, that the leading input of the second
differential is guaranteed, if the matrix ${\rm A}_2$ is positively
defined and diverges, i.e., if \be A_2^{-1}\rightarrow 0.\ee In this
`hard'\, classical regime one obtains, that $\dot p_{\mu}=0$, i.e.
$p_{\mu}= p_{0\,\mu}=p_{\mu}(t_0)$, and that \be\label{sol}
\dot{x}^{\mu}= -\,\frac{2}{m}\,p_0^{\mu}.\ee Note, that the `hard'\,
classical dynamics does not depend on $\beta$, so in this
approximation the temperature regime $\beta=\beta(t)$ does not
affect on trajectory. Also, from Eq. (\ref{sol}) it follows, that
the thermodynamical time $t$ is proportional to the proper time of
the particle under consideration. Of course, one can rescale the
energy operator $E$ to make these time variables coinciding. In
doing so, we mean, that $p_{0\mu}p_0^{\mu}=-m^2c^2$, i.e. that the
mass parameter $m$ is related to the initial momentum value.

Thus, in the case of highly-resonance distribution of the
probability density, effectively one has the single time in the
theory dynamics. However, in the general situation, where one deals
with nontrivial `probability stream'\, not only on the single world
line, but in the whole space-time, one comes back to the classical
dynamics with the sufficiently double time structure. Of course, it
is possible to split this `probability stream'\, into congruence of
the (world) lines defined by the tangent moment vector field
$p_{\mu}$. For the resulting world line structure,
$x^{\mu}=x^{\mu}(t)$ for the any fixed temperature regime
$\beta=\beta(t)$. These world lines can be curved: it is easy to
understand using the discussed extremal trajectories in the general
case of ${\rm A}_2^{-1}\neq 0$, see Eqs. (\ref{rel14}),
(\ref{rel15}). It is interesting to note, that the curved world
lines originate in the free theory: actually, the Hamiltonian taken
is the coordinate-independent and quadratic in respect to the
moments.

Such the dynamical properties make the classical theory under
consideration really close to the General Relativity, where the free
motion is realized on the curved world lines too \cite{gr}. We hope
to develop this analogy as far as it possible `to derive gravity
from quantum mechanics'.\, This program will not only follow to
reformulation of the pure geometrical General Relativity in the
probability terms of the quantum mechanics: our general scheme also
contains a temperature and decay dynamical variables. Thus, it seems
possible to develop a modification of the General Relativity with
different temperature regimes and arrow of time (at least for the
physically important part of them). Such generalized theoretical
framework will be really useful for cosmology and black hole physics
\cite{bh-f}--\cite{cosm}. Actually, in these geometrically based
disciplines the correct appearance of the arrow of time is one of
the most waited, important and not realized property.

\vskip 10mm \noindent {\large \bf Acknowledgements}

\vskip 3mm \noindent We would like to thank prof. B.S. Ishkhanov for
many discussions and private talks which were really useful for us
during this work preparation. One of the authors (O.V.K) was
supported by grant ${\rm MD \,\, 3623.\, 2006.\, 2}$.

\end{document}